# Loop Antennas for Use On/Off Ground Planes

John J. Borchardt, *Member, IEEE*

*Abstract*— Many applications benefit from the ability of an RFID tag to operate both on and off a conducting ground plane. This paper presents an electrically small loop antenna at 433 MHz that passively maintains its free-space tune and match when located a certain distance away from a large conducting ground plane. The design achieves this using a single radiation mechanism (that of a loop) in both environments without the use of a ground plane or EBG/AMC structure. An equivalent circuit model is developed that explains the dual-environment behavior and shows that the geometry balances inductive and capacitive parasitics introduced by the ground plane such that the free-space loop reactance, and thus resonant frequency, does not change. A design equation for balancing the inductive and capacitive parasitic effects is derived. Finally, experimental data showing the design eliminates ground plane detuning in practice is presented. The design is suitable for active, "hard" RFID tag applications.

## I. INTRODUCTION

Many reported UHF RFID antenna designs are suitable for use on a conducting ground plane [1], and many applications further benefit from the ability to operate *both* near to and away from a ground plane. This feature is known by various names in the literature: "platform-tolerant," "dual-environment," "on/off metal," etc. Many such designs have been reported, for example [2]-[32] and [41]-[44]. Designs based on dipoles ([2]-[11]), PIFAs ([12]-[21]) and patches [22]-[32] are common. Some designs, such as [4], [29], [44], operate as one type of radiator in free-space and as another type when near ground.

Most antenna designs surveyed [2]-[44] have a height about 1% of the operating wavelength $\lambda_0$; only five ([3], [11], [25], [29] & [41]) have heights greater than $0.03\lambda_0$. Low profile designs are suitable for "soft" RFID tags for small items where flexibility, low height profile and low cost are primary concerns. RFID of large items such as shipping containers, railroad cars and industrial machinery often requires longer read range—and thus higher antenna performance. In such applications, "hard" RFID tags with rigid plastic housings, higher height profile (often about $0.03\lambda_0$ [45]-[49]) and higher cost are often used; the design presented herein is suitable for such hard tag applications.

Although loop antennas are common in inductively coupled, near-field RFID applications, they are relatively uncommon in the far-field UHF RFID literature for either on-metal only [38]-[40] or dual-environment applications [41]-[44]. However, a popular design comprised of two PIFAs whose radiating edges face each other [16]-[21] (which are platform-tolerant) and [33]-[35] (which are not platform-tolerant) resembles a loop antenna in some regards, as do "folded patch" designs [22] and [23], the "high impedance unit cell" design of [36] and the "looped bowtie" of [37]. Because electrically small loop antennas interact with the magnetic field, they are a natural choice for use against a conducting ground plane—where incident tangential magnetic fields are doubled [41]. Unfortunately, loop antennas are known to detune when brought near conducting ground planes [50], hindering their use in dual-environment applications.

This paper presents a simple loop antenna that balances inductive and capacitive parasitics introduced by a ground plane such that the free-space tune and match change little when the antenna is located a fixed distance away from a ground plane. In practical hard tags (e.g., [45]-[49]), this distance is already naturally fixed by a dielectric radome. This work contributes to the RFID literature by 1) giving a simple antenna geometry that performs well in both free-space and near a conducting plane, 2) developing a simple equivalent circuit model that explains the observed dual-environment behavior and 3) deriving a general design equation from the equivalent circuit that allows others to easily scale the design.

The design herein is about $0.03\lambda_0$ in height—similar to [45]-[49]—and is suitable for 50 Ω active hard tag applications in the 433 MHz ISM band. However, the technique presented is readily adapted to passive 915 MHz soft tags. For example, because loop antennas are *inductive*, it is easy to adjust the matching network used herein for conjugate match to capacitive input passive RFID chips—less matching capacitance is used than in the 50 Ω case [42]. Moreover, the discrete matching capacitors used in this work may be easily integrated into a two-layer etched inlay as in [42]. Finally, the loop conductor may be etched on a flexible inlay that is folded around a foam support as in [15], [22], [23] & [40]. This work extends [41] by giving 1) a heuristic explanation of the observed platform-tolerant phenomenon, 2) a quantitatively accurate equivalent circuit model and design equation for achieving platform-tolerance and 3) experimental data showing that the platform-tolerant antenna works in practice.

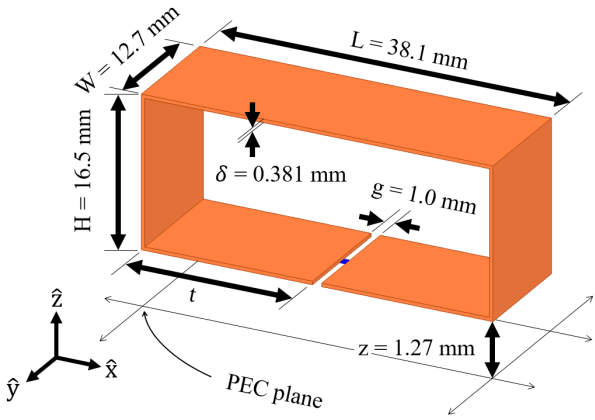

*Figure 1. Copper ($\sigma$=58MS/m) loop antenna above an infinite metal ground plane at z=0; the loop is fed at the g=1.0 mm gap. t is set in Sections II & III. An L-section matching network is used to tune and match the loop for a free-space (z=$\infty$) environment (see Fig. 2).*

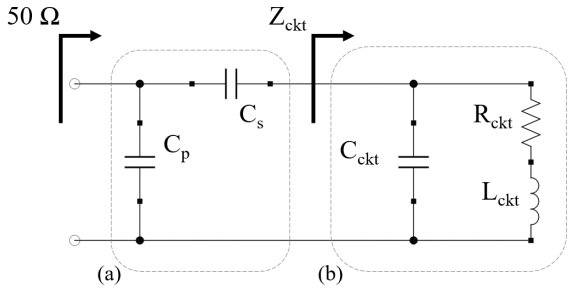

*Figure 2. (a) L-section matching network [53], and (b) equivalent circuit of the loop in free-space [54].*

## II. Loop Antenna Detuning Near a Ground Plane

The severity and nature of the detuning problem for small loop antennas used in both free-space and near ground planes is demonstrated via full-wave calculation of the Fig. 1 geometry in free-space (z=$\infty$) with t=18.55 mm using Ansys HFSS [51]. Note that basic image theory considerations [52] dictate the loop orientation shown is the most profitable for radiation near a ground plane.

The resulting full-wave reactance, $X_{fw}$, is shown in Fig. 3(a). The free-space loop is matched to 50 $\Omega$ at 433 MHz with an L-section matching network [53] consisting of $C_s$=4.19 pF series and $C_p$=149 pF shunt capacitors, as shown in Fig. 2(a); the resulting reflection coefficient is shown in Fig. 3(b). When this *same loop geometry* is simulated z=1.27 mm away from a ground plane and this *same matching network* is applied, the resonance shifts 2% to 424 MHz, also shown in Fig. 3(b). In this state, the mismatch loss at 433 MHz is greater than 20 dB. However, near 424 MHz the return loss is 10 dB, resulting in only 0.5 dB mismatch loss; if this resonance shift can be eliminated, good performance can be achieved on/off metal.

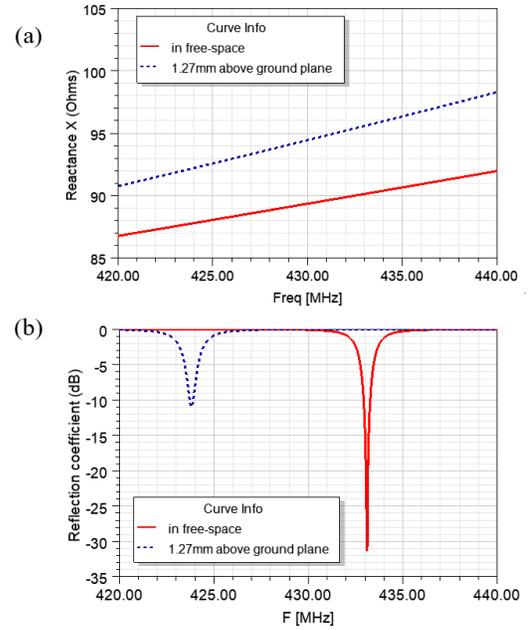

*Figure 3. (a) Full-wave reactance, $X_{fw}$, of the Fig. 1 loop (with no matching network) with t=18.55mm in free-space and z=1.27 mm above a ground plane. (b) Resulting reflection coefficient when the antenna is matched with identical networks shown in Fig. 2(a). The resonance shifts only 2%, but causes 20 dB mismatch loss at 433 MHz.*

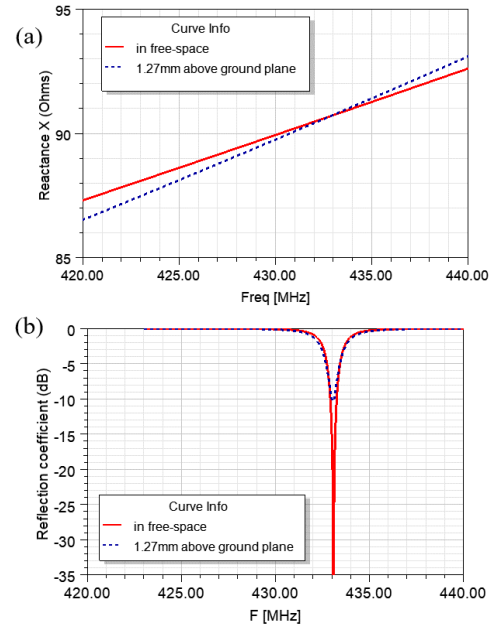

*Figure 4. (a) Full-wave reactance, $X_{fw}$, of the Fig. 1 loop (i.e., with no matching network) with t=7.2 mm for free-space and z=1.27 mm above a ground plane. The reactance in both environments is equal at 433 MHz. (b) Resulting reflection coefficient when the antenna is matched with identical networks of Fig. 2(a), showing negligible detuning between the two environments.*

## III. A Loop Antenna With Stable Tuning

The Fig. 1 loop is altered such that t=7.2 mm. The full-wave loop reactance, $X_{fw}$, is calculated for the loop in free-space and z=1.27 mm away from the ground plane. As seen in Fig. 4(a), $X_{fw}$= 90.7 $\Omega$ at 433 MHz for *both* environments. The free-space antenna may be matched to 50 Ω at 433 MHz with $C_s$=4.17 pF series and $C_p$=148 pF shunt capacitors as shown in Fig. 2(a). Because the reactance does not change between the two environments, the matched resonant frequency shifts negligibly, as shown in Fig. 4(b). There is a modest change in the full-wave loop resistance, $R_{fw}$, between the environments (about 0.122 Ω in free-space and 0.230 Ω when $z$=1.27 mm away from the ground plane). Despite this, the return loss remains acceptable—about 10 dB.

The full-wave calculated radiation efficiency of the loop is 57% in free-space and 76% when $z$=1.27 mm away from the ground plane; matching network losses may be considered separately. Because the loop circumference is only 16% of the wavelength, the current distribution is substantially uniform. Consequently, the radiation pattern approximates a torroid in free-space and a half-torroid when the loop is located close to a large ground plane; currents and patterns are shown in Fig. 5.

## IV. EQUIVALENT CIRCUIT

A simple circuit model explains the observed stable-tuning phenomenon and yields a design equation achieving this at an arbitrary frequency below the loop self-resonance. The free-space loop can be represented by the parallel-resonant equivalent circuit of Fig. 2(b) [54]. Circuit element values can be selected such that the equivalent circuit impedance $Z_{ckt}$ matches the full-wave model impedance $Z_{fw}$ at the operating frequency $\omega_c$ as follows. First, the self-resonant frequency $\omega_0$ of the loop in free-space is calculated via the full-wave model. Next, the full-wave model is solved at the operation frequency $\omega_c$ yielding $Z_{fw} = R_{fw} + j\,\omega_c\,L_{fw}$. The equivalent circuit resistance $R_{ckt}$ and inductance $L_{ckt}$ are then chosen using the following approximations derived from basic circuit analysis of the Fig. 2(b) equivalent circuit:

$$R_{ckt} \sim R_{fw}(1-\omega_c^2/\omega_0^2)^2, \quad (1)$$

$$L_{ckt} \sim L_{fw}(1-\omega_c^2/\omega_0^2). \quad (2)$$

Finally, $C_{ckt}$ is determined according to:

$$C_{ckt} = 1/(\omega_0^2 L_{ckt}). \quad (3)$$

For the Fig. 1 geometry with $t$=7.2 mm, full-wave simulation indicates the free-space loop resonant frequency $\omega_0 = 2\pi 1.2$ GHz and $Z_{fw} = 0.122 + j90.7$ Ω at $\omega_c = 2\pi 433$ MHz. By (1), (2) and (3), $R_{ckt}$= 0.092 Ω, $L_{ckt}$= 29.0 nH and $C_{ckt}$= 0.606 pF.

When the free-space loop is brought near a ground plane, the primary perturbation is magnetic flux coupling between the loop and its ground plane image, shown in Fig. 6 as mutual inductance $M$. The capacitance $C_1$ in Fig. 6 represents the parasitic capacitance between the feed node and its ground-plane image; with respect to the feed, $C_1$ appears in series with the larger capacitance $C_2$ of the remaining loop bottom face. Thus, the effective value $C_{eq} = (C_1 C_2)/(C_1 + C_2)$ is approximately $C_1$ when $t \ll L$[1]. Without $C_{eq}$, there is no frequency below self-resonance where the reactance of the free-space equivalent circuit (shown in Fig. 2(b)) and that of the near-ground equivalent circuit (shown in Fig. 7(a)) are equal.

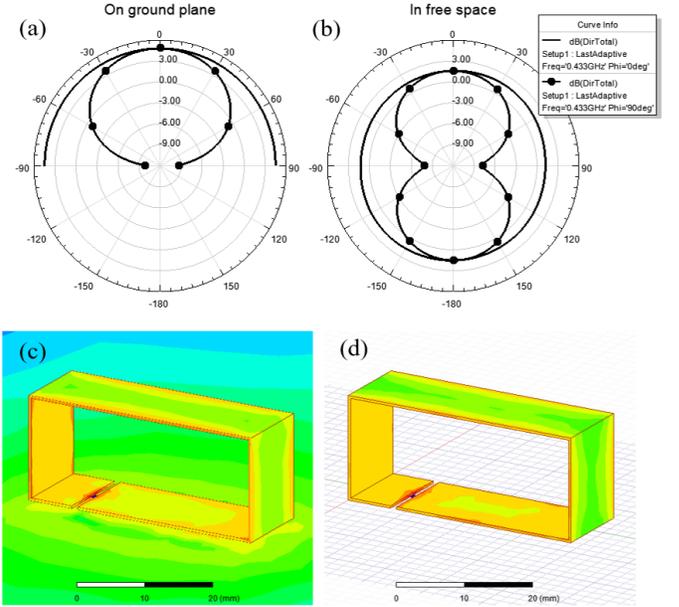

Figure 5. Directivity (a) and (b) and current magnitude (c) and (d) for the antenna of Fig. 1 with $t$=7.2 mm near a large ground plane (a) and (c), and in free-space (b) and (d). The patterns approximate an ideal toroid in free space and a half-toroid when near the ground plane. The predominant loop current vector direction is in the plane of the loop; the color scale is +40 dBA/m to -40 dBA/m.

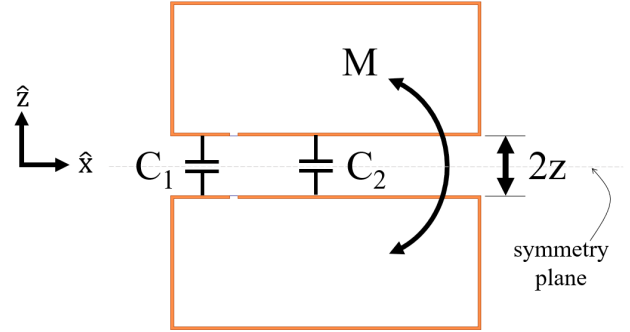

Figure 6. Loop antenna with $t$=7.2 mm (top) and its ground-plane image (bottom) separated by a distance $2z$. The mutual inductance $M$ between the loops and two parasitic capacitances are indicated. Because $C_1$ is smaller than $C_2$ and the two appear in series, the total series capacitance $C_{eq}$ is close to $C_1$.

With $t$=7.2 mm and $z$=1.27 mm, a simple estimate of $C_{eq} \sim C_1$ using the parallel plate capacitor formula is 0.325 pF. The mutual inductance $M$ is calculated from the full-wave two-port s-parameters of the mirrored Fig. 6 geometry at 10 MHz as 3.42 nH. The transformer coefficient $k$ is thus $M/L_{ckt}$= 0.118.

The equivalent circuit of the loop near a ground plane *without a matching circuit* is shown in Fig. 7(a). Because image theory dictates the image source has 180° phase [52], the equivalent circuit operates in the "odd" (anti-symmetric) mode only [55]. As illustrated in Fig. 7(b), odd mode operation allows for a greatly simplified ground-plane equivalent circuit, shown in Fig. 7(c) that is topologically equivalent to the free-space circuit of Fig. 2(b). Using basic circuit theory and a high $Q$ approximation, we find the reactance of the Fig. 7(c) equivalent circuit is equal to that of the Fig. 2(b) free-space equivalent circuit at the operation frequency $\omega_c$ when:

$$\omega_c^2 \sim \frac{1}{2}\left(\frac{k}{1-k}\right)\left(\frac{1}{C_{eq}L_{ckt}}\right) \quad (4)$$

---

[1] The series arrangement of $C_1$ and $C_2$ implies that $C_{eq}$ is maximum when the feed gap is centered along the loop length $L$ (i.e., when $t \sim L/2$); due to symmetry, either increasing or decreasing $t$ from the centered position will reduce $C_{eq}$ from the maximum value.

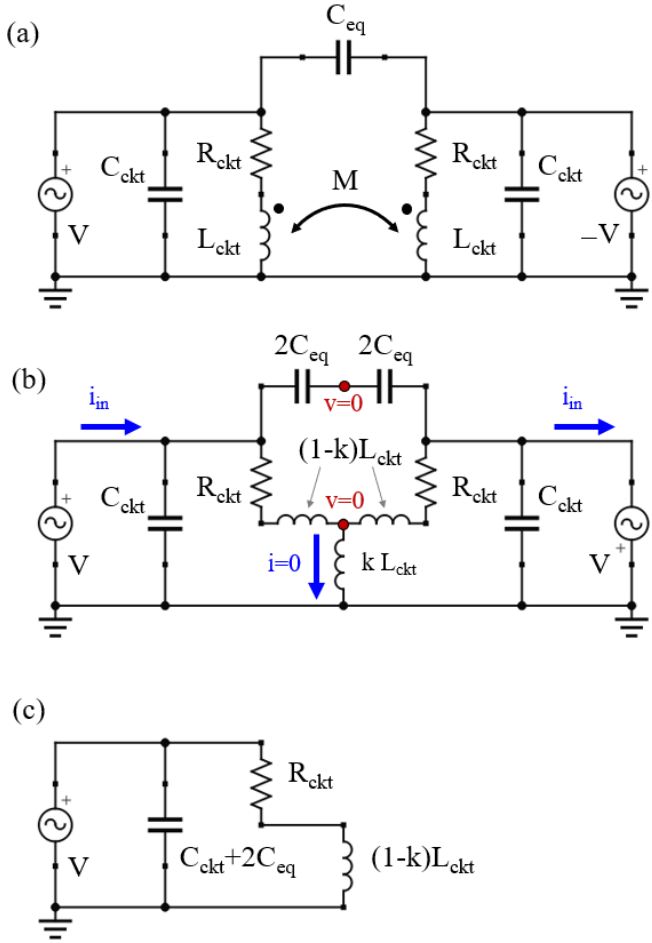

Figure 7. (a) Equivalent circuit of the loop near a ground plane, and (b) modified equivalent circuit. Image theory dictates an antiphase source image and thus odd (anti-symmetric) mode operation, resulting in zero shunt inductor current and zero potential at the nodes indicated; thus, these nodes may be grounded, yielding the simplified circuit of (c). No matching circuit is shown in (a), (b) or (c).

This relation predicts that the equal-reactance frequency decreases as $C_{eq}$ is increased (e.g., by increasing $t$ from 7.2 mm) and as the loop inductance is increased (e.g., by decreasing $W$)—as is observed in both the full-wave model as well as experimentally.

For $\omega_c = 2\pi 433$ MHz, $k=0.118$, and $L_{ckt}=29.0$ nH, (4) indicates $C_{eq}$ must be 0.312 pF, in good agreement with the parallel-plate capacitance estimate. Circuit model results are shown in Fig. 8.

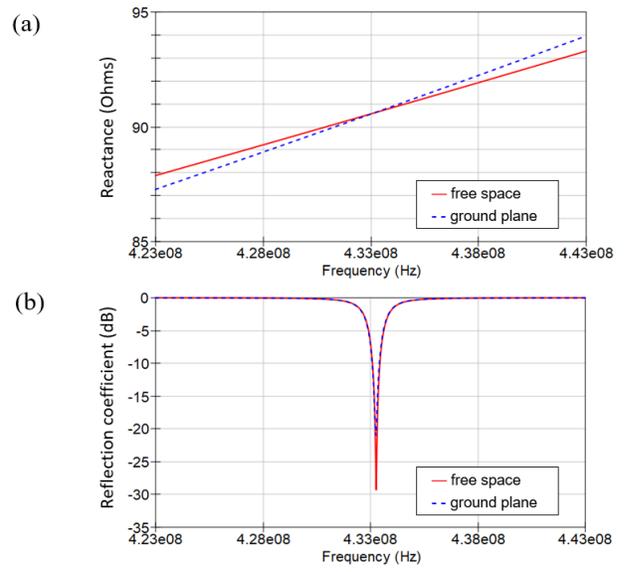

Figure 8. Circuit model results for free-space (i.e., Fig. 2(b)) and z=1.27 mm above ground plane (i.e., Fig. 7(a)), with $C_{eq}$ tuned to 0.305 pF (the value predicted by (4) is 0.312 pF); (a) unmatched loop reactance and (b) 50 Ω reflection coefficient with $C_s$= 4.17 pF and $C_p$=148 pF L-section matching network. The circuit models show equal reactance at 433 MHz and resulting zero resonance shift.

From (4), we see that only certain combinations of $k$, $C_{eq}$ and $L_{ckt}$ yield equal reactance at the operating frequency. Thus, for a given loop, stable tuning occurs only for a particular ground plane separation $z$. In practical hard tags (e.g. [45]-[49]), this distance is already fixed by the radome. Because $C_{eq}$ is influenced by the radome dielectric, its effect should be accounted for via full-wave simulation. Regardless, the platform-tolerant principles presented above remain valid.

## V. IMPLEMENTATION AND TESTING

The Fig. 1 loop geometry with $t=7.2$ mm was implemented with copper sheet and an FR-4 printed circuit board (PCB) as shown in Fig. 9. The PCB hosts the matching network and a U.FL coaxial connector. The loop is fixed $z=1.27$ mm away from a 16" round metal ground plane using a nylon screw and nylon washers. The reflection coefficient was measured with a vector network analyzer (VNA) whose cable was arranged to minimally couple to the antenna loop (i.e., the loop formed by the measurement cable and ground plane was orthogonal to the antenna loop) as shown in Fig. 9. RF chip capacitors in the L-section matching network were adjusted until the input impedance was close to 50 Ω at 433 MHz as seen in Fig. 10. Note that $C_s$ and $C_p$ were each implemented via multiple chip capacitors in parallel to reduce losses and allow for fine-tuning. A hemispherical Wheeler cap measurement [56] of the matched loop on ground yielded 40% radiation efficiency (including matching losses).

Next, the VNA cable was disconnected from the antenna and a VNA-connected, sub-resonant magnetic field probe (Beehive Electronics model 100A) was held within a few centimeters of the loop antenna. Normally, such a probe reflects nearly all RF power incident from the VNA. However, at the loop antenna resonant frequency, a small amount of RF power is coupled from the probe; a small dip in the VNA-measured reflection coefficient is evident, indicating the antenna resonance. (This is a classic technique known as "grid dip" [57].) In this case, the probe-indicated resonance was about 2 MHz from that of the initial cabled VNA measurement; this implies the cable did not significantly impact the cabled VNA measurement. Note that in practice, removing a small amount of PCB feed node copper with a razor can tune $C_{eq}$ to finely adjust the detuning characteristic in accord with (4).

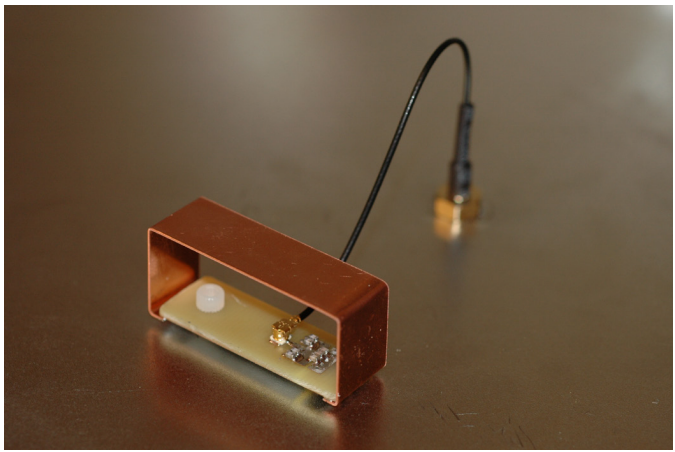

*Figure 9. As-built loop antenna with t=7.2 mm located on a ground plane with the VNA measurement cable connected. The bottom face of the loop is spaced 1.27 mm off the ground plane with a nylon screw and nylon washers. The feed gap in the PCB copper is visible in the ground plane reflection of the loop underside.*

Next, the ground plane was removed, and the magnetic field probe measurement repeated. The indicated resonance shifted less than 1 MHz, demonstrating that the loop antenna maintains stable tune between the two environments as designed; this data is shown in Fig. 10.

Finally, a miniature, battery-powered swept-frequency RF source with approximately constant-available power in 50 Ω was set inside the loop. The RF signal transmitted by the loop was monitored with a polarization-matched, sub-resonant monopole probe connected to a spectrum analyzer located about 60 cm away from the loop antenna in the $\hat{z}$ direction. This was done with the loop in three ground plane configurations: 1) $z=1.27$ mm away from the ground plane, 2) $z=75$ mm away from the ground plane, and 3) with no ground plane; results are presented in Fig. 11. When $z=75$ mm, full-wave modeling shows the loop impedance is close to that in a true free-space environment, however, the radiation pattern remains similar to that of the $z=1.27$ mm case (i.e., it is approximately a half-toroid). Little difference in the received peak amplitude and frequency was observed between configurations 1 and 2. When the ground plane is removed entirely, the radiation pattern changes significantly—now a full toroid covering $4\pi$ steradians—and the signal is expected to decrease 3 dB due to this effect. Moreover, the full-wave calculation predicts 1.25 dB decrease in radiation efficiency when moving from $z=1.27$ mm away from ground plane to free space. The measured signal decreased 4.5 dB, with little frequency shift. This confirms the antenna works as designed; we estimate the change in mismatch

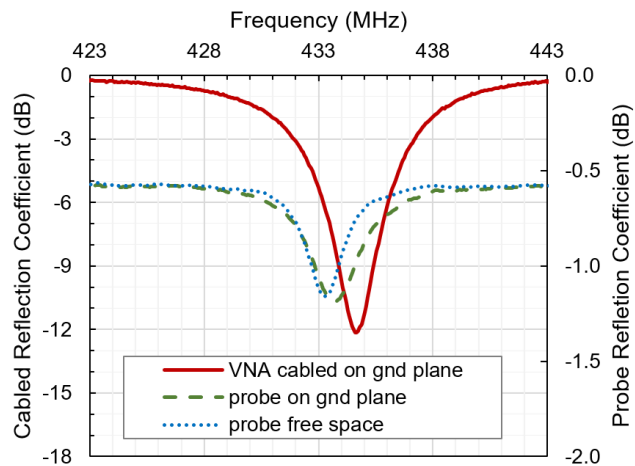

*Figure 10. Cabled VNA reflection coefficient of the loop z=1.27 mm away from the ground plane, as well as "grid dip" near-field probe measurements on and off the ground plane. All three resonances are very close to each other, indicating that the measurement cable is not perturbing the measurement and that the antenna has stable tune between the ground plane and free-space environments.*

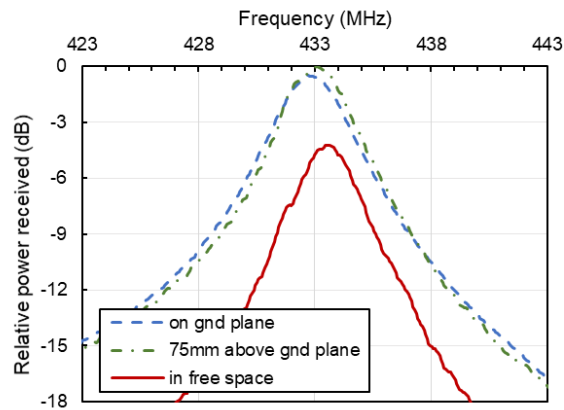

*Figure 11. Experimental power received from the transmitting loop antenna with t=7.2 mm in three environments demonstrating minimal frequency shift and limited change in peak amplitude. Accounting for expected changes in radiation pattern and radiation efficiency between the ground plane and free-space environments, we estimate the change in mismatch loss is about 0.25 dB.*

loss at the feed between free-space and ground plane environments is about 0.25 dB in this case.

## VI. CONCLUSION

This paper presents a general design principle for preventing detuning when electrically small loop antennas are used both in free space and a fixed distance away from a ground plane. The approach balances the inductive and capacitive parasitics introduced by the ground plane such that the resonant frequency remains unchanged. The design may be easily adapted to commercial passive UHF RFID applications via the design equation (4), flexible inlay fabrication and straightforward adjustment and implementation of the matching network to give a complex conjugate match to passive RFID integrated circuits.

ACKNOWLEDGMENT

The author gratefully acknowledges the contributions of Steve Dunlap, Craig Bennett, Leonard Dixon, Stephanie Otts, Cory Ottesen, Luke Feldner, Dylan Crocker, Thomas Roth and Jeff Williams to this work. Some aspects of this work are the subject of a US Patent application [58].